\documentclass[12pt, final, twocolumn, twoside, romanappendices]{IEEEtran}

\usepackage{amsmath,amssymb}

\usepackage[bookmarks,colorlinks]{hyperref} 
\hypersetup{colorlinks,citecolor= red,filecolor= blue,linkcolor= blue,urlcolor=blue}

\usepackage{graphicx}%
\usepackage[usenames,dvipsnames]{xcolor}

\usepackage{tikz,pgfplots}

\usepackage[printonlyused,withpage]{acronym}
\usepackage{cite} 
\usepackage{color}
\usepackage{comment} 

\usepackage{WINS-LatexInclusion/Preambles/winsnotation} 
\usepackage[level=3]{WINS-LatexInclusion/Preambles/wgroup_message}  
\acrodef{BPC}[BPC]{broadening participation in computing}
\acrodef{BPE}[BPE]{broadening participation in engineering}
\acrodef{CISE}[CISE]{Computer and Information Science and Engineering}

\acrodef{PI}[PI]{Principal Investigator}
\acrodef{KI}[KI]{Key Investigator}
\acrodef{RO}[RO]{research objective}
\acrodef{RA}[RA]{research activity}
\acrodefplural{RA}[RA]{research activities}

\acrodef{STEM}[STEM]{science, technology, engineering, and mathematics}

\acrodef{FY}[FY]{fiscal year}
\acrodef{KPI}[KPI]{key performance indicator}

\acrodef{AFOSR}[AFOSR]{Air Force Office of Scientific Research}
\acrodef{ARO}[ARO]{Army Research Office}
\acrodef{DARPA}[DARPA]{Defense Advanced Research Projects Agency}
\acrodef{DOD}[DoD]{Department of Defense}
\acrodef{ONR}[ONR]{Office of Naval Research}
\acrodef{ISR}[ISR]{intelligence, surveillance and reconnaissance}
\acrodef{SA}[SA]{situational awareness}
\acrodef{MDMP}[MDMP]{military decision making process}
\acrodef{COA}[COA]{course of action}
\acrodef{OODA}[OODA]{observe, orient, decide, and act}

\acrodef{AA}[AeroAstro]{Aeronautics and Astronautics}
\acrodef{EECS}[EECS]{Electrical Engineering \& Computer Science}

\acrodef{SCC}[SCC]{Schwarzman College of Computing}
\acrodef{MIT}[MIT]{Massachusetts Institute of Technology}
\acrodef{IDSS}[IDSS]{Institute for Data, Systems, and Society}
\acrodef{ISN}[ISN]{Institute for Soldier Nanotechnologies}
\acrodef{LIDS}[LIDS]{Laboratory for Information and Decisions Systems}
\acrodef{WINSLAB}[WINS\,Lab]{Wireless Information and Network Sciences Laboratory}

\acrodef{OMEFAC}[OMEFAC]{Office of Minority Education's Faculty Advisory Committee}
\acrodef{MAES}[MAES]{Mexican American Engineers and Scientists}
\acrodef{HSSP}[HSSP]{high school studies program}

\acrodef{UROP}[UROP]{Undergraduate Research Opportunities Program}
\acrodef{MSRP}[MSRP]{MIT Summer Research Program}
\acrodef{URM}[URM]{underrepresented and minority}

\acrodef{IIT}[IIT]{Indian Institute of Technology}
\acrodef{SIT}[SIT]{Stevens Institute of Technology}
\acrodef{UA}[UA]{University of Arizona}

\acrodef{LIDAR}[LIDAR]{light detection and ranging}
\acrodef{NSTC}[NSTC]{National Science \& Technology Council}
\acrodef{NLA}[NLA]{noiseless linear amplification}
\acrodef{QIS}[QIS]{quantum information science}
\acrodef{QN}[QN]{quantum network}
\acrodef{QLS}[QLS]{quantum localization and synchronization}
\acrodef{QLSN}[{\it q}LSN]{quantum localization and synchronization network}
\acrodef{QNS}[QNS]{quantum network science}
\acrodef{QSD}[QSD]{quantum state discrimination}
\acrodef{RADAR}[RADAR]{radio detection and ranging}

\acrodef{AC}[AC]{actor-critic}
\acrodef{ACK}[ACK]{acknowledge}
\acrodef{AE}[AE]{angle estimate}
\acrodef{AI}[AI]{angle information}
\acrodef{AII}[AII]{angle information intensity}
\acrodef{AIS}[AIS]{automatic identification system}
\acrodef{AL}[AL]{angle likelihood}
\acrodef{ALU}[ALU]{arithmetic logic unit}
\acrodef{ANN}[ANN]{artificial neural network}
\acrodef{AOA}[AOA]{angle-of-arrival}
\acrodef{API}[API]{application programming interface}
\acrodef{ASIC}[ASIC]{application-specific integrated circuit}
\acrodef{AWGN}[AWGN]{additive white Gaussian noise}
\acrodef{AWS}[AWS]{Amazon Web Services}
\acrodef{BAC}[BAC]{Bayesian actor-critic}
\acrodef{BC}[BC]{belief condensation}
\acrodef{BCF}[BCF]{belief condensation filter}
\acrodef{BLE}[BLE]{Bluetooth Low Energy}
\acrodef{BP}[BP]{belief propagation}
\acrodef{BPZF}[BPZF]{band-pass zonal filter}
\acrodef{BTB}[BTB]{Bellini-Tartara bound}
\acrodef{BTZZB}[BTZZB]{Bellini-Tartara Ziv-Zakai bound}
\acrodef{CCDF}[CCDF]{complementary cumulative distribution function}
\acrodef{CDF}[CDF]{cumulative distribution function}
\acrodef{CE}[CE]{cross-entropy}
\acrodef{CEO}[CEO]{counting error outage}
\acrodef{CESS}[CE-SS]{cross-entropy SS}
\acrodef{CF}[CF]{characteristic function}
\acrodef{CFR}[CFR]{channel frequency response}
\acrodef{CIR}[CIR]{channel impulse response}
\acrodef{CNS}[CNS]{classical network science}
\acrodef{CR}[CR]{channel response}
\acrodef{CRB}[CRB]{Cram\'{e}r-Rao bound}
\acrodef{CRLB}[CRLB]{Cram\'{e}r-Rao lower bound}
\acrodef{CSD}[CSD]{Cauchy-Schwarz divergence}
\acrodef{CSI}[CSI]{channel state information}
\acrodef{CV}[CV]{continuous variable}
\acrodef{DA}[DA]{data-association}
\acrodef{DE}[DE]{distance estimate}
\acrodef{DFE}[DFE]{decision feedback equalizer}
\acrodef{DINS}[DINS]{decentralized inference in networked systems}
\acrodef{DNI}[DNI]{decentralized network inference}
\acrodef{DP}[DP]{dynamic program}
\acrodef{DPa}[DP]{direct path}
\acrodef{DPOMDP}[Dec-POMDP]{cecentralized-partially observed Markov decision process}
\acrodef{DRT}[DRT]{distance ratio test}
\acrodef{DV}[DV]{discrete variable}
\acrodef{EBIT}[EBIT]{entanglement bit}
\acrodef{ED}[ED]{energy detector}
\acrodef{EED}[EED]{Engineering Every Day}
\acrodef{EFIM}[EFIM]{equivalent Fisher information matrix}
\acrodef{EKF}[EKF]{extended Kalman filter}
\acrodef{EIRP}[EIRP]{equivalent isotropically radiated power}
\acrodef{ESD}[ESD]{energy-based soft-decision}
\acrodef{ESPRIT}[ESPRIT]{estimation of signal parameters via rotational invariant techniques}
\acrodef{FCC}[FCC]{Federal Communications Commission}
\acrodef{FG}[FG]{factor graph}
\acrodef{FII}[FII]{Fisher information inequality}
\acrodef{FIM}[FIM]{Fisher information matrix}
\acrodef{FL}[FL]{feature likelihood}
\acrodef{FP}[FP]{feature potential}
\acrodef{FSC}[FSC]{finite state controller}
\acrodef{FW}[FW]{Fisher-Wald}
\acrodef{FY}[FY]{fiscal year}
\acrodef{GDOP}[GDOP]{geometric dilution of precision}
\acrodef{GLMB}[GLMB]{generalized labeled multi-Bernoulli}
\acrodef{GLRT}[GLRT]{generalized likelihood ratio test}
\acrodef{GMMLMB}[G-MM-LMB]{Gaussian MM-LMB}
\acrodef{GNSS}[GNSS]{global navigation satellite system}
\acrodef{GP}[GP]{Gaussian process}
\acrodef{GPS}[GPS]{Global Positioning System}
\acrodef{HCA}[HCA]{heterogeneous computing architecture}
\acrodef{HDSA}[HDSA]{high-definition situation-aware}
\acrodef{HDP}[HDP]{hierarchical Dirichlet process}
\acrodef{HI}[HI]{hard information}
\acrodef{HMM}[HMM]{hidden Markov model}
\acrodef{IID}[IID]{independent, identically distributed}
\acrodef{IMU}[IMU]{inertial measurement unit}
\acrodef{INDFT}[INDFT]{inverse non-uniform discrete Fourier transform}
\acrodef{INR}[INR]{interference-to-noise ratio}
\acrodef{IOT}[IoT]{Internet-of-Things}
\acrodef{IR-UWB}[IR-UWB]{impulse radio UWB}
\acrodef{IRPR}[iRPR]{infinite regionalized policy representation}
\acrodef{JBSF}[JBSF]{jump back and search forward}
\acrodef{KDE}[KDE]{kernel density estimation}
\acrodef{KF}[KF]{Kalman filter}
\acrodef{KL}[KL]{Kullback-Leibler}
\acrodef{KLD}[KLD]{Kullback–Leibler divergence}
\acrodef{LBP}[LBP]{loopy belief propagation}
\acrodef{LEM}[LEM]{Laplacian eigen-map}
\acrodef{LEO}[LEO]{localization error outage}
\acrodef{LIDAR}[LIDAR]{light detection and ranging}
\acrodef{LMB}[LMB]{labeled multi-Bernoulli}
\acrodef{LMS}[LMS]{least means square}
\acrodef{LOCC}[LOCC]{local operations and classical communication}
\acrodef{LOS}[LOS]{line-of-sight}
\acrodef{LOT}[LoT]{Localization-of-Things}
\acrodef{LQG}[LQG]{linear-quadratic-Gaussian}
\acrodef{LRT}[LRT]{likelihood ratio test}
\acrodef{LRFS}[LRFS]{Labeled random finite set}
\acrodef{LS}[LS]{least squares}
\acrodef{LSE}[LSE]{line spectral estimation}
\acrodef{MAC}[MAC]{medium access control}
\acrodef{MAP}[MAP]{maximum a posteriori}
\acrodef{MBS}[MBS]{maximum bin search}
\acrodef{MC}[MC]{Monte Carlo}
\acrodef{MDD}[MDD]{minimum distance distribution}
\acrodef{MDP}[MDP]{Markov decision process}
\acrodef{MF}[MF]{matched filter}
\acrodef{MHT}[MHT]{multi-hypothesis tracking}
\acrodef{MIMO}[MIMO]{multiple-input multiple-output}
\acrodef{ML}[ML]{maximum likelihood}
\acrodef{MLE}[MLE]{maximum likelihood estimation}
\acrodef{MMSE}[MMSE]{minimum-mean-square-error}
\acrodef{MMLMB}[MM-LMB]{merged-measurement LMB}
\acrodef{MOT}[MOT]{multi-object tracking}
\acrodef{MOU}[MOU]{measurement origin uncertainty}
\acrodef{MP}[MP]{map potential}
\acrodef{MSE}[MSE]{mean-square error}
\acrodef{MUI}[MUI]{multi-user interference}
\acrodef{MUSIC}[MUSIC]{multiple signal classification}
\acrodef{NBI}[NBI]{narrowband interference}
\acrodef{NGF}[NGF]{NII generation function}
\acrodef{NIC}[NIC]{network inference encoding}
\acrodef{NISQ}[NISQ]{Noisy Intermediate-Scale Quantum}
\acrodef{NII}[NII]{network inference information}
\acrodef{NLN}[NLN]{network localization and navigation}
\acrodef{NLT}[NLT]{network localization and tracking}
\acrodef{NLOS}[NLOS]{non-line-of-sight}
\acrodef{NSF}[NSF]{National Science Foundation}
\acrodef{OOT}[OoT]{ocean-of-things}
\acrodef{OOTCESS}[OoT-CESS]{OoT-CESS}
\acrodef{OP}[OP]{outage probability}
\acrodef{OSPA}[OSPA]{optimum subpattern assignment}
\acrodef{P-Max}[P-Max]{$P$-Max}  
\acrodef{PAR}[PAR]{probabilistic association rule}
\acrodef{PCA}[PCA]{principal component analysis}
\acrodef{PDF}[PDF]{probability distribution function}
\acrodef{PDP}[PDP]{power delay profile}
\acrodef{PEB}[PEB]{position error bound}
\acrodef{PF}[PF]{physical features}
\acrodef{PHD}[PHD]{probability hypothesis density}
\acrodef{PMF}[PMF]{probability mass function}
\acrodef{POCS}[POCS]{projection onto convex sets}
\acrodef{POMDP}[POMDP]{partially observed Markov decision process}
\acrodef{PPM}[PPM]{pulse position modulation}
\acrodef{PPP}[PPP]{Poisson point process}
\acrodef{QUA}[QuaDRiGa]{QUAsi Deterministic RadIo channel GenerAtor}
\acrodef{RAT}[RAT]{radio access technology}
\acrodef{RCS}[RCS]{radar cross section}
\acrodef{RF}[RF]{radiofrequency}
\acrodef{RFID}[RFID]{radio frequency identification}
\acrodef{RFS}[RFS]{random finite set}
\acrodef{RI}[RI]{range information}
\acrodef{RII}[RII]{range information intensity}
\acrodef{RL}[RL]{reinforcement learning}
\acrodef{RLi}[RL]{range likelihood}
\acrodef{RLS}[R-LS]{range-based least squares}
\acrodef{RLMC}[RL-MC]{reinforcement learning Monte Carlo}
\acrodef{RM}[RM]{resource management}
\acrodef{RMS}[RMS]{root mean square}
\acrodef{RMSE}[RMSE]{root-mean-square error}
\acrodef{ROI}[ROI]{region of interest}
\acrodef{RPR}[RPR]{regionalized policy representation}
\acrodef{RRC}[RRC]{root raised cosine}
\acrodef{RSS}[RSS]{received signal strength}
\acrodef{RTT}[RTT]{round-trip time}
\acrodef{RV}[RV]{random variable}
\acrodef{SBS}[SBS]{serial backward search}
\acrodef{SBSMC}[SBSMC]{serial backward search for multiple clusters}
\acrodef{SC}[SC]{soft constraint}
\acrodef{SDE}[SDE]{stochastic differential equation}
\acrodef{SDN}[SDN]{software defined network}
\acrodef{SK}[SK]{soft knowledge}
\acrodef{SI}[SI]{soft information}
\acrodef{SII}[SII]{speed information intensity}
\acrodef{SIR}[SIR]{signal-to-interference ratio}
\acrodef{SLAM}[SLAM]{simultaneous localization and mapping}
\acrodef{SMC}[SMC]{sequential monte carlo}
\acrodef{SNR}[SNR]{signal-to-noise ratio}
\acrodef{SINR}[SINR]{signal-to-interference-plus-noise ratio}
\acrodef{SO}[SO]{soft observation}
\acrodef{SPAWN}[SPAWN]{sum-product algorithm over a wireless network}
\acrodef{SPEB}[SPEB]{squared position error bound}
\acrodef{SR}[SR]{sensor radar}
\acrodef{SS}[SS]{sensor selection}
\acrodef{SSCH}[SSh]{sensor scheduling} 
\acrodef{SVE}[SVE]{single-value estimate}
\acrodef{TBD}[TBD]{track before detect}
\acrodef{TCS}[TCS]{threshold crossing search}
\acrodef{TD}[TD]{temporal difference}
\acrodef{TDOA}[TDOA]{time difference-of-arrival}
\acrodef{TH}[TH]{time-hopping}
\acrodef{TNR}[TNR]{threshold-to-noise ratio}
\acrodef{TOA}[TOA]{time-of-arrival}
\acrodef{TOF}[TOF]{time-of-flight}
\acrodef{TSD}[TSD]{threshold-based soft-decision}
\acrodef{TWS}[TWS]{track while scan}
\acrodef{UAV}[UAV]{unmanned aerial vehicles}
\acrodef{UKF}[UKF]{unscendent Kalman filter}
\acrodef{UML}[UML]{unsupervised machine learning}
\acrodef{UWB}[UWB]{ultra-wideband}
\acrodef{VM}[VM]{von Mises}
\acrodef{VNA}[VNA]{vector network analyzer}
\acrodef{WAF}[WAF]{wall attenuation factor}
\acrodef{WBI}[WBI]{wideband interference}
\acrodef{WED}[WED]{wall extra delay}
\acrodef{WLS}[WLS]{weighted least squares}
\acrodef{WPAN}[WPAN]{wireless personal area network}
\acrodef{WSN}[WSN]{wireless sensor network}
\acrodef{WWB}[WWB]{Weiss-Weinstein bound}
\acrodef{XGL}[xGL]{next generation localization}
\acrodef{ZZB}[ZZB]{Ziv-Zakai bound}
\acrodef{ZZLB}[ZZLB]{Ziv-Zakai lower bound}

\definecolor{BLUE}{rgb}{0,0,1}

\DeclareMathAlphabet{\foo}{U}{tx-cal}{m}{n}

\newcommand{\hsa}[1] {h_{\text{s},\text{a}}}

\newcommand{\srxb}[1] {r_{\text{s-b},i}(t)}
\newcommand{\srxref}[1]{r_{\text{ref},i}(t)}
\newcommand{\srxrem}[1]{r_{\text{rem},i}(t)}

\newcommand{\thetaBi}[1]{\boldsymbol \theta_{B_i}}

\newcommand{\toar}[1]{\hat{\tau}_{i,0}}

\definecolor{myred}{rgb}{1,0.27,0}
\definecolor{mygreen}{rgb}{0.20, 0.8, 0.2}
\definecolor{myblue}{rgb}{0, 0, 0.8}
\definecolor{myorange}{RGB}{255, 178, 102}
\definecolor{mymagenta}{rgb}{0.78, 0.08, 0.52}
\definecolor{mycyan}{rgb}{0, 0.74, 1}

\usepackage[yyyymmdd,hhmmss]{datetime} 
\newdateformat{monthyeardate}{\monthname[\THEMONTH] \THEDAY, \THEYEAR} 
\usepackage[pagewise,mathlines,displaymath]{lineno} 

\usepackage{balance}

\newcommand{\paperTitle}{Packet Routing for the Quantum Internet}

\pgfplotsset{compat=1.18}
\begin{document} 

\twocolumn

\title{\paperTitle}


\author{
	\vspace{0.2cm}
	
   Robert Malaney\\
    \thanks{
    }    
    
    \thanks{
    
 The author is with the School of Electrical Engineering and Telecommunications, University of New South Wales, Sydney, Australia. email: r.malaney@unsw.edu.au

	}		
}

\maketitle 

\begin{abstract}
We present a new design for quantum packet routing within the  emerging Quantum Internet, highlighting how a little-used feature of  Internet Protocol Version~6 (IPv6), namely Extension Headers, can  lead to  a significant amount of quantumness within the IP layer.
Taking a minimalist approach to alterations of established standards, we outline the changes required in order for quantum teleportation, quantum routing, and superpositions of these processes to be enabled. 
Relative to other proposals for routing within the Quantum Internet, the architecture we propose enables a wider range of  outcomes allowed by quantum mechanics. We do not claim any optimally in our design, but rather a pathway to invoke new quantum routing outcomes via small additions to the current IPv6.

\end{abstract}


\acresetall		

\section{Introduction}\label{sec:intro}
\IEEEPARstart{P}{acket} routing in the emerging Quantum Internet  can occur via quantum processes that are simply impossible on the Classical Internet. Foremost among these are quantum teleportation, and quantum routing - the quantum superposition of routes through the network.\footnote{In many works, the term `quantum routing' is taken to simply mean the distribution of quantum entanglement through a network using single-path routing; here we take this term to mean the full superposition of routing paths allowed for by quantum mechanics.} In this work, we introduce within the context of the Open Systems Interconnection (OSI) model~\cite{osi}, a modified Internet Protocol (IP) packet format that seamlessly allows quantum teleportation and quantum routing, as well as superpositions of both quantum processes.

The packet switching procedures that underpin routing in the Classical Internet were devised around the 1970's. In search of a means to withstand significant interruptions in communication networks, packet switching without centralized control was proposed~\cite{internetv1}. Subsequently, the Classical Internet was born and expanded to encompass the World Wide Web, a sophisticated  communication system that went well beyond what anyone  originally imagined. Routing through the IP remains a critical component of this Web~\cite{web3}.

As was the case for the Classical Internet, we predict that the Quantum Internet will evolve and lead to new communication scenarios that are not currently imagined. It is therefore important that the operational protocols that are being developed for this new Internet  accommodate all possibilities allowed by quantum mechanics. 

The question we attempt to answer here is  simple. \textit{If we were} to adopt a layered approach based on the OSI Internet design, what minimum additions would be required to accommodate quantum teleportation, quantum routing, and superpositions thereof? This is a pragmatic question faced by current network operators who wish to slowly migrate into quantum networking, whilst being backward compatible with their classical infrastructure.
We answer our question by building on  the spectacularly successful IP packet format used in classical networks, adopting as a starting point IPv6~\cite{rfc8200}.  
\section {Related Work}
It is important to clarify that we do not claim that our  approach is  the optimal way forward for quantum packet routing. Indeed, counter arguments have been made as to why any traditional layered approach based on Classical Internet design should \textit{not} be followed~\cite{callno1,   cal8, caleffi1, durno1, callno2, illi}. In~\cite{callno1}, the issue of scalability was first raised as an argument against maintaining a layered approach for quantum networks. In~\cite{cal8}, the notion of native quantum routing was introduced in which the packet header information becomes quantum, which, when coupled with an entanglement-defined controller, allows scalable quantum control operations~\cite{caleffi1}.
In~\cite{durno1},
a resource-centric task-based scheme was proposed where a node that initiates the service requirement creates a distributed workflow.
The work of~\cite{callno2} drew attention to the global coordination required by a resource-centric scheme, which makes implementation challenging, and instead proposed a new architecture centered on packets with an in-band control containing the service information. In the  architecture of~\cite{callno2}, a pathway to scalability was offered - as the packet traverses the network, successive nodes construct local actions only. In~\cite{illi}, the notion of an entanglement-based switching fabric within a router was presented, leading to a hardware-agnostic solution.

Notwithstanding the above comments, other workers have considered a layered approach to quantum networking, \cite{y1,y2,y3,y4} being among many notable examples. In \cite{y1}, a new 4-layer quantum network stack was offered;
while in \cite{y2} a system design was offered for a network of quantum repeaters within a layered approach.
In \cite{y3}, a link layer protocol for quantum networks was offered within a layered approach; 
and in \cite{y4} a quantum data plane protocol was proposed, providing a building block for a layered quantum network.
Other works specifically focused on packet switching architectures within a layered approach~\cite{18p1,18p2,18p3,18p4}. 
In~\cite{18p1}, a specific frame structure was offered, while in~\cite{18p2} Quantum Key Distribution (QKD) within a packet switching framework was studied.
In~\cite{18p3}, the concept of `quantum wrapper networking' was introduced for control and management, with~\cite{18p4} providing an experimental demonstration of this concept.
Some previous studies entirely focused on the management of the entanglement distribution, a critical feature of future quantum networks (but a feature not required for all communication scenarios). Rather than any discussion on interaction between other layers, many of these studies largely described the functionality of the link layer~\cite{link1, link2}.   

However, none of the above-mentioned works focused on the quantum-based routing possibilities we  consider here, especially within the context of IPv6.  The architecture we propose is the first attempt to include quantum routing and superpositions of this process with other quantum processes within a layered approach.

Our study is timely: the experimental implementation of a quantum packet transfer based on the time synchronization of quantum and classical data (a  model that we assume for direct transmissions) has recently
 been carried out~\cite{laserpack};
and technical expositions of the quantum mechanics we utilize in our study are established, namely, quantum teleportation~\cite{tel1, tel2}, quantum routing~\cite{Oi,Gisin, Lemr2013, wenbo1} and superpositions of quantum processes in general~\cite{allproc}. The reader is referred directly to these references (and references therein) for technical details on the assumed quantum processes.

Although we will utilize IPv6 Extension Headers to invoke new  routing layer features in the quantum domain,  similar ideas have previously been utilized in the classical-only domain. A good example of this is the work of~\cite{class33} in which the Extension Headers are used to enhance data sovereignty at the routing layer. Security issues raised by reading (and dropping) IPv6 Extension Headers are discussed in~\cite{class34}. The reader is referred to~\cite{class33,class34} for a more extensive explanation, than that offered here, of the structure and interplay between the Base Header and the Extension Headers within IPv6.

\section{System Model}\label{sec:STQcom}
\begin{figure}
	\centering
	\includegraphics[width=.95\linewidth]{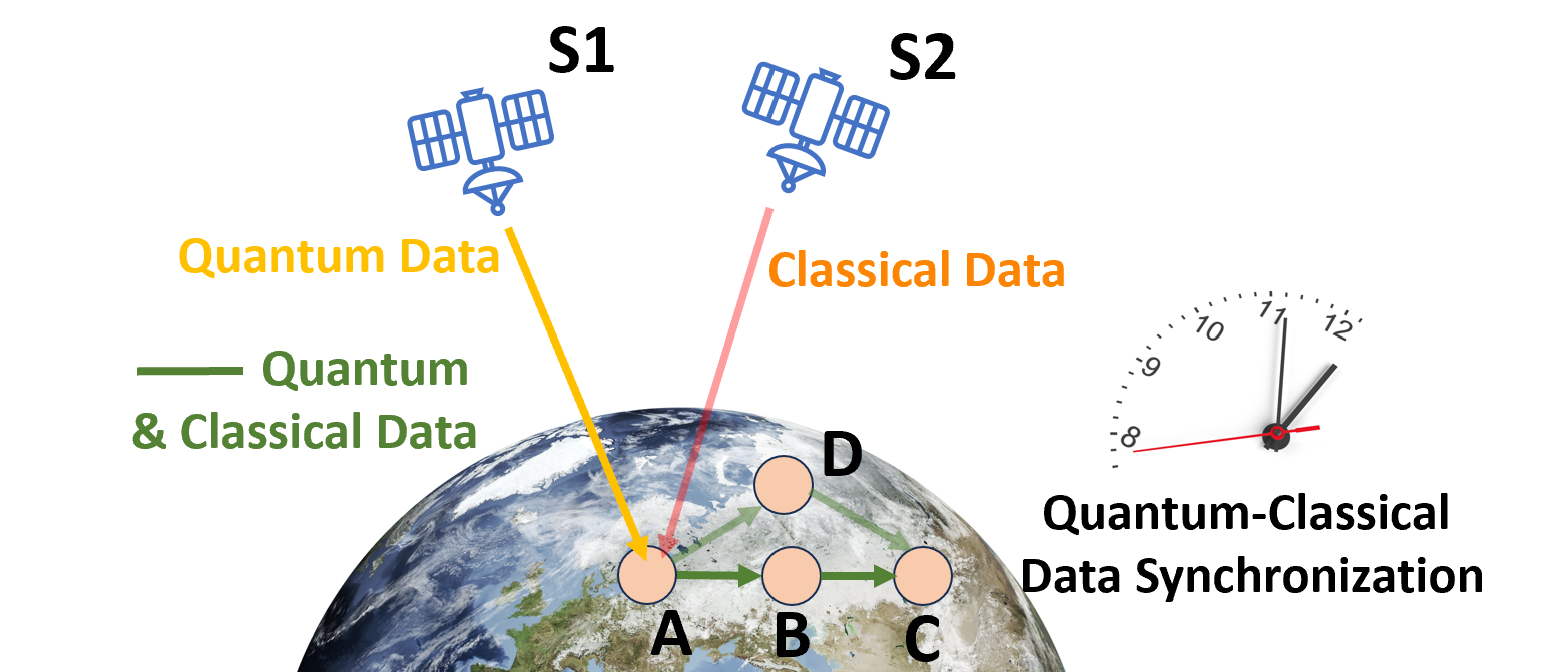}
	\caption{\textbf{System Model.} A typical system set up where IP routing will be required. Satellite S1 wishes to transmit quantum information to node C via nodes A and B. Here a second satellite S2 sends classical data (header information) synchronized to reach node A just ahead of the quantum information from S1 (systems where S1 also sends the classical information are also possible).  Node D may be used, as quantum mechanics allows for a superposition of paths ABC and ADC. The network is assumed perfectly synchronized. }
	\label{fig:network3}
\end{figure}

To provide clarity in our exposition, we will adopt the system model shown in Fig.\ref{fig:network3}, where a satellite S1 sends a quantum state to a router (node) C on Earth, but is passed through routers A and B to reach C. We note S1 and S2 need not be satellites in general, and our system can readily be adapted to fiber only system, or terrestrial free space systems, or some combination thereof. However, space-based quantum communications are now a reality, and we anticipate such communications to play a pivotal role in the emerging Quantum Internet~\cite{ref3}.

\subsection {Assumptions}
Our focus is on the IP header information used in the IP  layer (also called the network layer or simply layer 3) in our  quantum-enabled IP  stack, shown in Fig.\ref{stack}.\footnote{Multiplexing the classical and quantum signaling in other modes such as frequency or  polarization is also possible. For clarity we consider here only multiplexing in the time domain, given the recent experimental development of a chip designed specifically for that purpose~\cite{laserpack}.}
To simplify our exposition, the following assumptions are adopted in the system model.

1.	Quantum memory is available on all routers. Incoming quantum information can be stored and retrieved locally.\footnote{In principle, this requirement for quantum memory could be dropped, but this would require additional synchronization of network infrastructure and resources, however, we note that the required memory timescale that allows for IP processing  (including optical to electrical conversion) is already available via circulation in small optical fiber loops~\cite{optmem}.} The router itself manages the details of the ingress/egress of information to/from quantum memory.

2.	
Quantum  information is encoded in a single `pulse' of light, which could be a Discrete Variable (DV) single photon state or a multi-photon Continuous Variable (CV) state.  Classical information is also encoded in (separate) light pulses.\footnote{The use of radio for the classical data could be readily accommodated.}
In direct transmissions, pulses of classical and quantum information are scheduled to arrive in the required order, so a router knows which pulse to store in quantum memory and which pulse to read classically. More specifically,  all information arrives at each router in sequence; classical header first, followed by payload (the quantum data). The combined classical and quantum information is called a quantum packet.\footnote{Here, we will not add any trailer after the quantum payload to remain faithful  to the IPv6 standard~\cite{rfc8200}, however, given the complexity of quantum networks, the use of a trailer may be useful in future applications.} 

3. For transmission via teleportation, an evolved version of the quantum state is transferred (teleported) via quantum measurements at the sender to a stored quantum state in the receiver's memory, which is then  manipulated  at the receiver based on classical data transmitted  by the sender. The classical data transmitted indicates which quantum state held by the receiver (the local memory address) is to be manipulated and how.\footnote{In port-based teleportation, no manipulation of the quantum state is required, the classical data indicate which port the receiver is to read the quantum state from~\cite{port}.} In teleportation, the quantum data indicated in Fig.\ref{stack} is replaced by the classical data required for the quantum state manipulation at the receiver.\footnote{The information held in this classical data requires thought when superpositions of quantum processes involving teleportation are used; see later discussion.}

4.	All classical communication is flawless. 

5.  A link layer protocol is in place and the link layer transfers quantum information via direct transmissions.\footnote{Transmission via teleportation at the link layer is possible, but care is needed when some superpositions are involved; see later discussion.}

6. A reliable entanglement distribution protocol is in place. We discuss this in more detail later.


\subsection{IPv6 Headers}
\begin{figure}
	\centering
\includegraphics[width=0.95\linewidth]{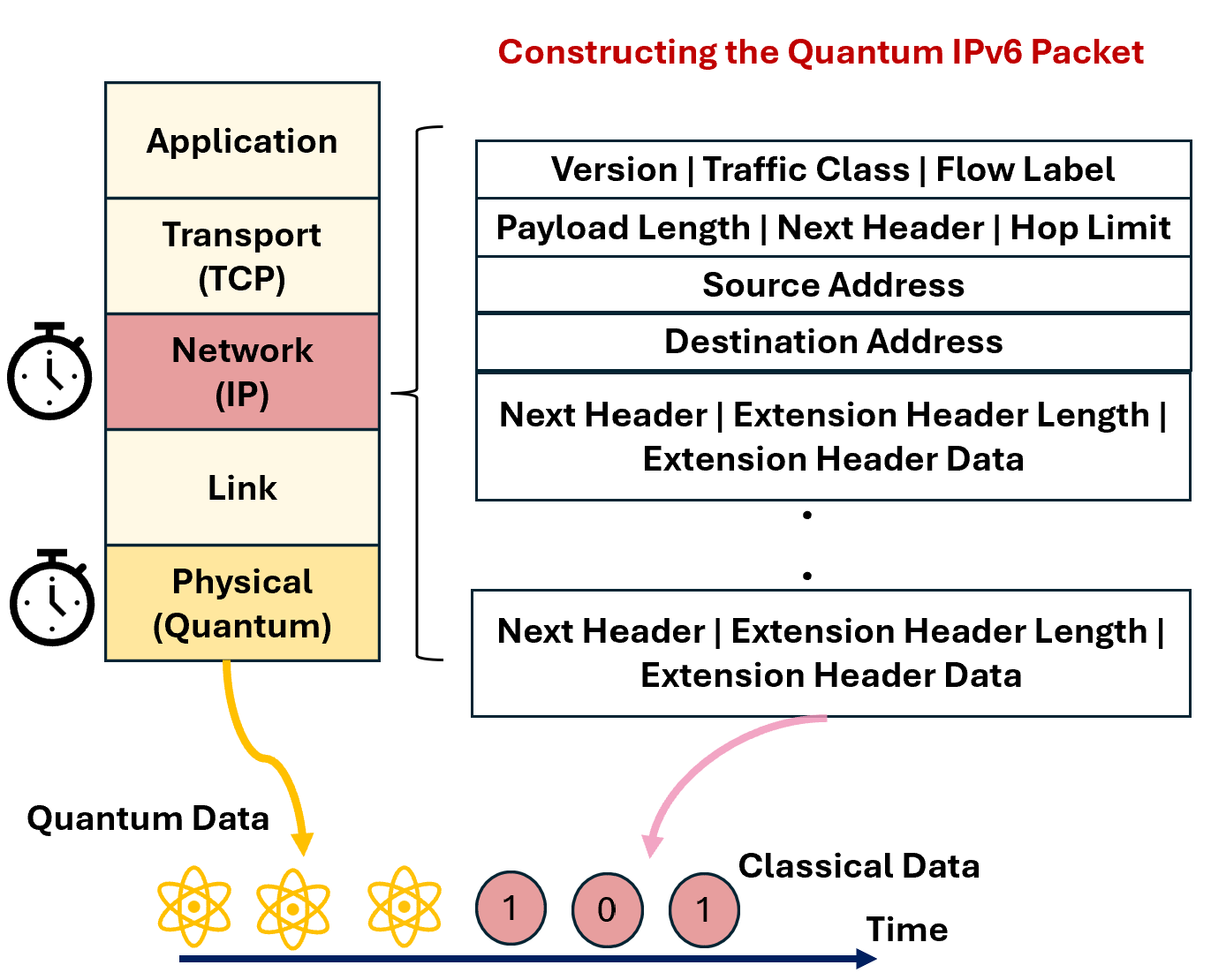}
	\caption{\textbf{The Quantum IP Stack.}  For direct transmission, classical data from the Base IPv6 Header and the Extension Headers are synchronized so that classical data arrives  at the next router's physical layer just prior to the quantum data. The quantum data is then stored in memory whilst the classical data is propagated up through the stack to the IP layer (teleportation involves a different transmission strategy). If the destination address corresponds to that router, the encapsulated TCP/UDP data is propagated to the Transport layer. Shown here are the standard fields in the IPv6 packet as per the standard~\cite{rfc8200}. Here, the top three layers of the OSI model are encapsulated into an application layer. This 5-layer stack is also referred to as the TCP/IP model. For teleportation, the quantum data is replaced by additional classical data.}
	\label{stack}
\end{figure}

The IPv6 Base Header field is the first four rows of data shown in Fig.\ref{stack}. This data is non-optional; it is always present in an IPv6 packet. The Base Header is of length 40 octets, followed by the Extension Headers. The combined information in the Base and Extension Headers is followed by data in the form of an upper layer Protocol Data Unit (PDU). The PDU contains the upper-layer protocol header and its payload, which could be a Transmission Control Packet (TCP) or a User Datagram Protocol (UDP) packet. in IPv6, the Extension Headers plus the upper-layer PDU are considered to be the payload of the packet.


 Most of the fields in the Base Header   are self-evident (see~\cite{rfc8200} for details), and here we discuss only some of them. One field refers to a quality-of-service paradigm referred to as Differentiated Services, and for our purposes we will assume that all packets have the same service. The Flow field is important as this allows different packets to be assigned to the same `flow.' This does not mean that IPv6 will force all packets in the same flow to pass through exactly the same path, but it allows processing such as re-sequencing. We will label all packets from a given source to a given destination with the same flow identifier. The Hop Limit field is always important in avoiding loops (see later discussion).
 
 The `Next Header' field (8 bits) in the Base Header indicates what happens next. If this points to a known protocol (such as 6 for TCP), the router knows that there are no Extension Headers to be read. If it points to a known (standardized) Extension Header field, the router knows it is to read that new header and act on it.  

 Given the assumptions embedded in our system model,  an IPv6  header consisting only of a Base Header (but with quantum payload)  should allow for a direct transfer of a quantum state through the network  via a single path from the source to the destination with only the slightest modifications to the network.
This raises the question; is there any circumstance that an  IPv6  header containing only a Base Header could NOT easily accommodate the routing of quantum information?  The answer to this question is clear: yes. The superposition of routing paths is one clear example. 

That a  photon can simultaneously be on two different paths  brings a multitude of benefits that are not possible otherwise, including improved quantum information throughput~\cite{Gisin,Chiri}, two-way communication with a single particle~\cite{DelSanto}, teleportation using separable states~\cite{allproc},
accessing quantum information stored in superposition at different sites~\cite{qramo}, and the in-principle ability to communicate over infinite length communication channels~\cite{infin}. To access  these non-intuitive quantum advantages within the IPv6 framework, we require new Extension Headers.\footnote{As an aside, we note that the notion of addresses being in the form of a quantum state, specifically the supposition of addresses, is something required for quantum random access memory (QRAM), a feature likely required by future quantum computers~\cite{qramo, QRAM}. Attempts have already been made to implement QRAM in current noisy quantum computers~\cite{shi6}.}
 
 Extension Headers have a `Next Header' field (8 bits), a  `Header Extension Length' field (8 bits), and then a field to contain data (variable length). This latter field contains different data depending on the specific Extension Header, but is  related to information that allows specific instructions to be set (also called `Options' in some headers). The structure of these data, and any padding, must be defined for each Extension Header, indicating the length of fields and the mapping of these fields to specific instructions. 
 As noted above, the Extension Header configures its own Next Header field, and the chain continues until an Extension Header points to a known transport protocol like TCP or UDP.   

 Many Extension Headers in the classical-only domain have been standardized, despite being  unused by many current routers~\cite{class34}. These include the Hop-by-Hop Header (field value 0), 
 and the Routing Header (field value 43)~\cite{iana}. For this work, one of the most important headers is the Hop-by-Hop Extension Header. This header forces the instruction data   to be read by every router on the path, and this can be useful in aiding the router to behave in a non-classical manner. This header uses  Type–Length–value (TLV) coding to map the Option bits to specific instructions. The reader is referred to~\cite{rfc8200} for more details on Extension Header usage in the classical-only domain.

Some in the community consider IPv6 Extension Headers as the only `design flaw' of the original IPv6 protocol. Initially intended as vehicles for additional functionality (that would be discovered by future research) without requiring the introduction of an entirely new protocol, 25 years of usage have shown Extension Headers to be largely  of niche value. In fact, most routers on the Internet will simply ignore them. The situation has progressed to the point that in some online blogs, arguments for their deprecation are presented. However,  here we argue for continued usage of the IPv6 Extension Headers in the context of hybrid classical-quantum networks. Through them, many quantum-only aspects of information transfer can be salvaged from a slightly modified IPv6-enabled  network.

\section{The Quantum Extension Headers} 
There is no definitive pathway to implement the effects of quantum mechanics within IPv6. One pathway would be to define new Option fields within the Hop-by-Hop Extension Header. This modified header would then need to be standardized by the Internet Engineering Task Force (IETF) and registered with the Internet Assigned Number Authority (IANA). Similarly, the current Routing Extension Header could be modified and standardized. We should warn the would-be designer; however, the sequential order of current Extension Headers must be enforced as per the standard~\cite{rfc8200}. Beyond this, completely new `Quantum Extension' Headers could be proposed and standardized - and we follow this pathway here. These headers are to be all classical, with the quantum data communicated separately as previously outlined.

Our new headers would have a similar format to  Extension Headers in the classical domain;   the first field would be a Next Header (8 bits), with the second field being the Header Extension Length (8 bits). This would include the length of the Data field (variable length) and the number of pulses containing the quantum information so that a router knows when the current quantum packet ends. The Data field  would be where the new data (the instruction set or a mapping to an instruction set) would be designed. This instruction set could inform the router to carry out specific instructions through software or inform  the router to trigger some embedded hardware. In the following, we focus on the structure of this Data field, bearing in mind that this same structure could also be envisaged as new instruction sets within current Extension Headers.

In discussing new IPv6 Extension Headers for the quantum domain, we will provide some detailed specification for the first of these we specify,  highlighting only the major conceptual issues in the others as we progress through the paper.  Our aim here is to  highlight the broad role new IPv6 Extension Headers may be able to play in the emerging Quantum Internet, and any specific formatting we provide is to be taken as suggestive rather than definitive. In this vein, we do not categorically state the size of any fixed length field, or the structure and form of the padding and coding used to specify instruction sets or Options. \textit{This level of detail will require wide-spread participation of the community followed by a rigorous standardization process.}

It  becomes a matter of design in how to best construct a new Extension Header for IPv6 that encompasses as much non-classical behavior as possible. In the following any field should not be used in a circumstance where it makes no sense (e.g., the time to send field  in superpositions of quantum processes).
\subsection{Quantum Routing}
\begin{figure}
	\centering
	\includegraphics[width=.95\linewidth]{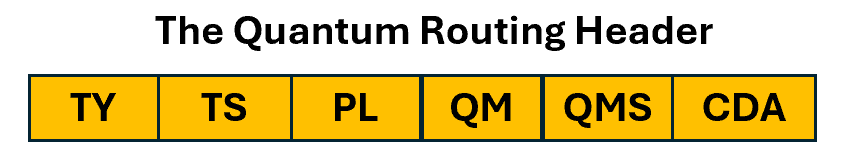}
	\caption{\textbf{Quantum Routing Header.} A proposed new IPv6 Extension Header that allows for various path superposition outcomes.}
	\label{Ipv6ExtOptions}
\end{figure}

The order in which these new headers appear will be important. As we  discuss later, the header that supports the superposition of paths in most circumstances should come before the header that supports teleportation. As such, we first discuss the former, referring to it as the `Quantum Routing' Header.
A suggested design for the Data field of this new header is shown in~Fig.~\ref{Ipv6ExtOptions}. We focus on the broad sub-fields that need defined, bearing in mind that the detailed structure and inclusion of any additional sub-fields are left to  standardization submission stages. Note that there are countless scenarios that a quantum routing header must encompass. So, design of such a header requires a minimum set of base rules that supersede any setting of any field. The most important of these is that no action by any router on any path is to destroy any requested superposition of paths.\footnote{An example of this is when a setting would lead to a router revealing `which-way' information on the paths. Any setting which leads to an unintended destruction of \textit{any} requested superposition (of paths or processes) is to be avoided - the designs outlined do not automatically guarantee maintenance of superposition for all setting combinations.}  The suggested sub-fields are as follows.

\textbf{Type (TY) Field:} This   field indicates the type of data in the quantum payload, DV (and its dimension), CV, or some hybrid combination. It can also stipulate whether the data are entangled or not (if known). We set  the Type setting to the number one for  a DV state of dimension two.\footnote{Single photon DV input states and corresponding Bell-State photon-pairs   as entanglement resources are likely to be widely used in the Quantum Internet. However,  new Extension Headers can cover any input state, and any entanglement resource state including a Two-Mode Squeezed Vacuum (TMSV) state and a hybrid DV-CV state.}

\begin{figure}
	\centering
	\includegraphics[width=.95\linewidth]{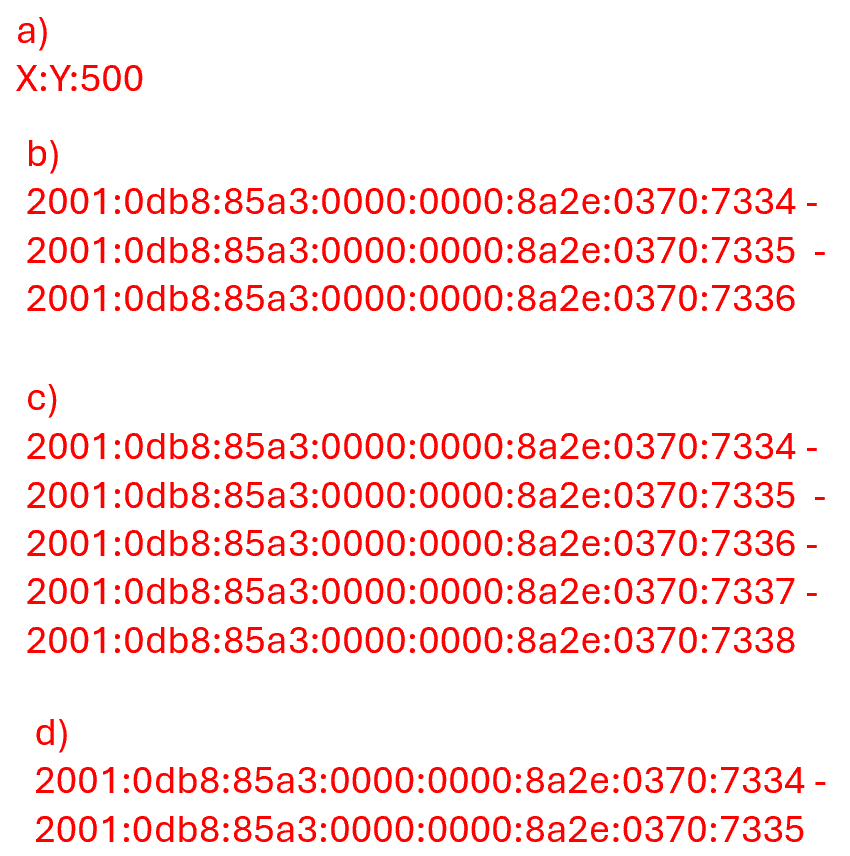}
	\caption{\textbf{Quantum Routing Format.} The IPv6 Quantum Routing format for various scenarios.}
	\label{packets}
\end{figure}

\textbf{Time to Send (TS) Field:}  A router uses this field to mark the time it received (X) and sent (Y) the quantum packet, plus an increment $\delta$. A non-zero $\delta$ indicates the time the next router should hold the quantum packet before sending. It is assumed that all routers on the network are synchronized with some global reference  (time units  picoseconds).
The format shown in Fig.\ref{packets}(a) is to be followed by the  router  sending the quantum packet.  The specific values shown  indicate that a current router received this packet at time X and sent the packet at time Y and expects that the next router will store the quantum information for 500 units of time before resending. All data by default in this field are initially set to -1,-1,-1, which means that a router does not  act on the field (and is not updated). If it is to be used, the initial values should be set to  $t$,~0,~$\delta$,  where $t$ is the current time. By default $\delta=0$ (the packet is to be sent as soon as possible).

\textbf{Path List (PL) Field:} This field lists all addresses where path splitting (superposition of paths) is to occur and then concatenates the next hop IP addresses. By default, it is set to zero and no splitting of paths is invoked. 
After splitting, all paths must be routed to the original destination IPv6  address. 
The format shown in Fig.\ref{packets}(b) is to be followed for this process.  
This example  would have a two-path splitting at the router …7334, with one path directed to router …7335 and the other path directed to router …7336. The format shown in Fig.\ref{packets}(c)  would have a four-path splitting at the router …7334, with the other two paths directed to router …7337 and the other additional path directed to router …7338. If the next hop router IP address has no listing in the local router table (or channel deemed invalid), the router itself is to pick the next hop IPv6 address - consistent with any constraint imposed by the routing algorithm (e.g. in a location-based algorithm the router is to pick the next router deemed closest to destination). Each router will be forced to inspect this PL field, which allows multiple splittings along the paths (even different types and numbers of splittings on the different paths). 

Note that in the splitting of paths, the classical header field is to be copied into all paths and sent separately to the quantum payload. It is only the quantum payload that is to be placed into a superposition of paths. At each router, the classical header information is propagated to the IP layer and acted upon. The quantum payload is meanwhile stored in local memory at the router. This creates a superposition of quantum memories and therefore a mechanism for invoking QRAM. 

\textbf{Quantum Multicast (QM) Field:} If this field is set to the address of some router (i.e non-zero),  the quantum state is to be  put into a superposition of all possible single-hop paths emanating from that router. We refer to this outcome as Quantum Multicast.\footnote{It is assumed that the first eight bits in the IPv6 addresses used are not fixed to ff00::/8 (binary 11111111) as would be the case for classical multicast within IPv6.} The format shown in Fig.\ref{packets}(d) is to be followed for this process.  This  specific setting would initiate a quantum multicast at router …7334 and also at router …7335. Again, all paths must lead to the original final destination. By default, the ingress path to a router is not to be included in the superposition (this is overridden by adding an asterisk at the end of an address).
The initial Hop Limit field in the Base Header should be set with particular thought for this setting (noting that it is likely decremented to different values in different paths).
This field supersedes the Path List field.

\textbf{Quantum Multicast Switch (QMS) Field:} By default, set to zero, but if set to 1, the Quantum Multicast field is ignored, and a quantum multicast is to take place at all routers on the paths. The final destination for all paths remains the original destination. 
The ingress path to a router is not included 
in the superposition. This field also supersedes the
Path List field.

\textbf{Change of Destination Address (CDA) Field:} This indicates if a change of final destination is to occur and can be set on any router. By default, this is set to zero (indicating that the original destination address in the IPv6 header remains untouched). If not set to zero, the new destination address is listed in this field and is to be inserted into the IPv6 Header. This allows for the different paths to now  have different destinations (i.e., have different header information).

The above fields accommodate most scenarios that could be envisaged for quantum routing.\footnote{Some of the combinations of these fields currently have no identified network purpose but are accommodated in case future research leads to new motivations.} Given the flexibility offered by IPv6, additional fields could be added. However, it is also possible to create entirely separate Extension Headers that compliment the above functionality, adding them into processes at the router via the Next Header field. Of course, here it is assumed that the target router  has the hardware functionality to invoke the required operations. Classical-only routers will not be affected by the passage of any Quantum  Header as they will be pre-programmed to simply ignore these Extension Headers.

\subsection{Teleportation}
We now discuss the Quantum  Header required for teleportation, which we refer to as the `Quantum Teleportation' header. 
Quantum networks will depend  at some level (i.e., not always) on the quantum entanglement distributed throughout the network. Many nodes will share entanglement, but every node will not share entanglement with every other node. This paradigm offers another quantum-only aspect of the network, teleportation. 
We delegate to the IP layer (or higher layers) decisions whether to teleport directly to nodes beyond the current one. The main reason for this is that, in general, requests for path superposition at certain nodes would need to take preference over teleportation direct to a receiving address. This is one reason why the Quantum Teleportation Header may need to be added to the Base IPv6 Header \textit{after} the Quantum Routing Header.\footnote{This is not an absolute requirement. A specific configuration for the superposition of routing and teleportation is that of Eq.~(7) of~\cite{allproc}, where two individual teleportation maps are invoked. Various combinations of the Quantum Extension Headers (including modifications to the Quantum Routing Header) can accommodate that configuration.}
Note, in the following, we assume that the link layer can intelligently adopt to information in this new header and override previous settings to transmit directly to the next router on a physical path. `Point-to-point' communication when teleportation is involved refers to communication via any viable quantum channel. Any required manipulation of entanglement may be made at other layers,\footnote{We use the phrase `other layers' to mean either the link layer, Application layer, or the TCP layer - or a combination thereof. It remains an open debate as to how exactly quantum networks will manage entanglement distribution (see later discussion).} but all routers will continuously update their routing tables, not only for classical communications as they currently do, but also for the availability of point-to-point quantum channels through teleportation. 

\begin{figure}
	\centering
	\includegraphics[width=.77\linewidth]{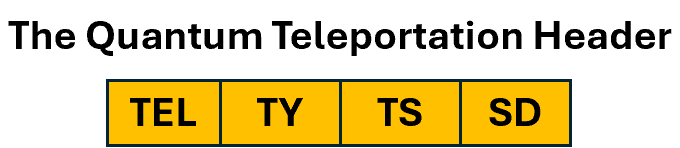}
	\caption{\textbf{Quantum Teleportation Header.} A proposed new IPv6 Extension Header that allows for various teleportation outcomes.}
	\label{Ipv6ExtOptions2}
\end{figure}

 A suggested format for this header is shown in Fig.\ref{Ipv6ExtOptions2}.
Again, we
focus on the broad inputs that likely need to be defined.

\textbf{The Teleportation (TEL) Field :}
The TEL field will have a fixed-bit number associated with it, allowing for mapping to a wide range of instruction sets. Most of these mappings are left open for future use, development, and standardization. However, we list a few sets of instructions that we believe will be of particular value and widely used. 
Note also that the Quantum Teleportation Header is copied and passed through all superpositions of paths (if any such paths are invoked by the Quantum Routing Header). The  quantum state to be teleported is the state taken directly from memory. Recall that memory now may be entangled with other memory addresses across the network.\footnote{Retrieving  memory from a superposition of addresses, most of which are not local to a specific router, in itself sets up a need for a new network protocol that allows such access. We do not discuss the details of that new access protocol, but it should be clear that this could also be built within the IPv6 Extension Header architecture and even in part built  using some of the new headers we are describing.} A few specific settings of the TEL field are now described.

\textbf{TEL=0} The default value is zero, which means that teleportation of the quantum state is to supersede all other instructions in the Quantum Routing Header if a `teleportation path' exists through the network. The router is to take the sender-receiver IP addresses from the Base IPv6 header and then make its best effort to teleport the quantum state in the minimum number of teleportation steps. If no teleportation is available along the path (and cannot be made available), then the network is to use direct transmission, if available, as required. 

{\textbf{TEL=1}:}
 This allows for specific teleportation transmissions rather than allowing the system to determine paths. For each pair of sender-receiver addresses input, the other layers again attempt to teleport in the minimum number of teleportation steps between those addressees. The pair of sender-receiver addresses to be input is provided in the header's Sub Data  field (defined below).  This setting supersedes instructions in the Quantum Routing Header. 

{\textbf{TEL=2}:} This is the same as stipulated above for the TEL=1 setting, but this time the information in the Quantum Routing Header will take precedence.

{\textbf{TEL=3}:} This forces the routers indicated to perform specific instructions regarding teleportation. Again, the instructions and routers requested to act are listed in the SD field. This generic setting allows for any possibility and is akin to a control message within an SDN type configuration. The information in the Data field can also request which information in different headers take precedence to each other, thereby offering a rich variation of possibilities.

\textbf{Type  (TY) Field:} This field is the same as that in the router header.

\textbf{Time to Send (TS) Field:} This field is the same as that in the router header.

\textbf{Sub Data (SD) Field:} This is where additional information and instructions related to the other header types is to be located.

Regarding the update of quantum teleportation channels in routing tables, we expect a process similar to that described for location-based routing in \cite{malaney5}.  This involves routers periodically announcing hello packets and constructing pathways based on these packets.  Individual routers in the network will be pre-programmed as to whether they will or will not participate in this process. If they are part of the process, the individual router may also set the value of how often they will participate (setting the period of their announcements). Upon receiving a request for teleportation to a specific destination, the other layers may respond by performing a sequence of entanglement swaps to establish a direct channel.  This process is managed according to the detailed entanglement distribution protocol adopted.

Routing via teleportation obviously represents one of the key new features that will make quantum networks truly distinct from their classical counterparts, and a feature that is likely to be widely utilized. In our process outlined, with the TEL field set to its default value, the network will make its best effort to transport the quantum state through the network using the minimum number of teleportation steps possible. It is possible that no teleportation steps are actually invoked (e.g., due to failures at the other layers in setting up  teleportation channels) and direct transmission proceeds or some combination of direct transmission and teleportation occurs.

There are other quantum processes  beyond quantum routing and teleportation that can also affect IP routing in the quantum context. One such process is the superposition of time orders (also commonly referred to as indefinite causal orders), whereby the temporal order of paths taken through a network is placed in superposition - a process known to enhance communication outcomes~\cite{casor}. We will not detail the header format required to invoke such processes within the IPv6 Extension Header format - suffice to say that it will follow similar structures to those outlined for quantum teleportation, particularly in the use of an equivalent to the TEL=3 field.

Note that it is also the case that combinations of existing classical IPv6 Extension Headers can be layered with the Quantum Extension 
Headers to form more complex routing behavior. As mentioned earlier, an example of this would be the classical Hop-by-Hop Extension Header, a means of forcing every router on a path to inspect the IP header. If this were to be implemented, the strict ordering rules of the classical headers need to be enforced. It is also possible that the classical  ICMP header  within IPv6 could be used for additional control features that could assist in the implementation of the new quantum aspects of the network.

\subsection{Superposition of Quantum Processes}
It is now well established that the superposition of quantum processes can give rise to enhanced communication outcomes. This includes  the superposition of quantum routing and quantum  teleportation~\cite{allproc}. Let us consider how such a superposition of different quantum processes can be enacted within the IPv6 framework.

\begin{figure}
	\centering
	\includegraphics[width=.95\linewidth]{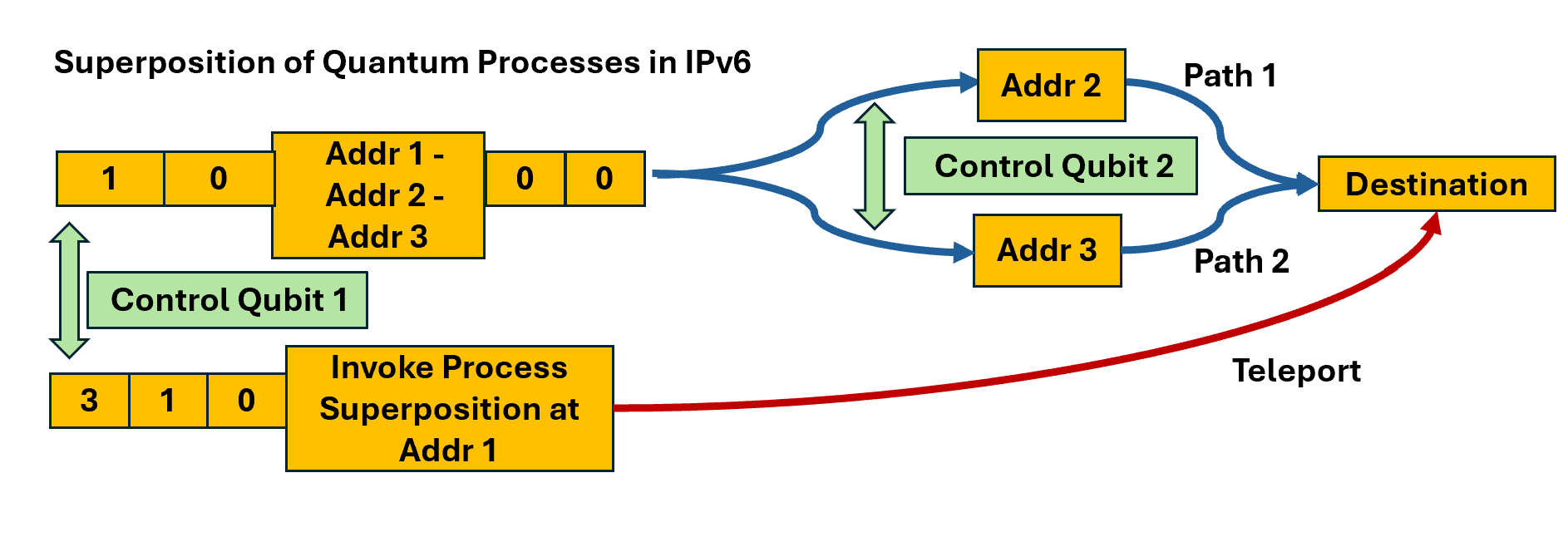}
	\caption{\textbf{Quantum Superposition of Quantum Processes.} One superposition of Quantum Routing and Quantum Teleportation where the hardware at a router located at Addr~1  invokes a superposition of the processes, triggered by Control Qubit 1. Here, the path superposition is triggered by Control Qubit 2 transmitting the quantum information to the destination in superposition via addresses Addr~1 and Addr~2. The information in the quantum Routing Header (top left) is copied and transmitted classically along the two paths in blue. The  information in the Teleportation Header (bottom left) is transmitted classically to the destination.  Different teleportation maps and control qubits could be added to this schematic so as to form different combinations (or orderings) of the superpositions. Not all combinations will have a specific quantum advantage in place. Technical details and quantum circuits that allow  superpositions with known quantum advantage are detailed elsewhere~\cite{allproc}.}
	\label{Ipv6ExtOptions3}
\end{figure}

Given the flexibility already available through the Quantum Routing Header and the Quantum Teleportation Header (with the TEL=3 setting), a superposition of both quantum processes could  be invoked using these new headers. One pathway for this would be to invoke new instructions set in the SDF of the Quantum Teleportation Header to instruct a particular router to  trigger the hardware that places the routing request in the previous Extension Router Header into a superposition with information in the  Quantum Teleportation Header fields.\footnote{Note, any measurement of a control qubit (either directly or indirectly (e.g., via Bell measurements) can collapse the superposition unless that classical information is stored (unread) in some ancilla state, an issue that does not necessarily hamper the application being pursued.} A schematic of such a process is shown in Fig.\ref{Ipv6ExtOptions3}. 

However, as mentioned earlier, other quantum processes beyond quantum routing and teleportation can be envisaged, in which we may also like to form superpositions.\footnote{For example,  the work of~\cite{sen25} a superposition of teleportation and classical communication is presented; also, superposition of a process with the previously mentioned process of indefinite causal orders is also possible.}. This suggests a better pathway would be the introduction of a `Quantum Superposition' Extension Header. We will not delve into details of such a header field, suffice to say that it will have many of the usual fields such as length of data,  but it will have a special `SUP' field, which indicates the processes to be put into superposition. This header will also have its own SDF which stipulates any additional information needed that was not available from the fields set in each process, or required clarifications on how exactly the superposition is to proceed (e.g., where and when such superpositions are to occur). A new header is likely a better way to handle the superposition of the quantum processes, as it allows future outcomes motivated by future research to be readily integrated into the network.

\section{Entanglement Distribution} 

We have left a substantial amount of `heavy lifting' within our architecture to a presumed operational entanglement distribution process within the network, especially in regard to teleportation requests. As discussed earlier, many works have discussed how this may be handled directly via sophisticated link layer protocols, and these may indeed prove ultimately successful. However, in the spirit of the IP approach we have adopted here, we prefer to ascribe to the link layer protocol the same responsibility assumed in the OSI model (transmissions between two nodes). We will assume the link layer adopts direct transmission for the transfer of the quantum information, but do note that this is not an absolute. Teleportation at the link layer is also possible, but care must be taken in the release of the classical outcome of the required quantum measurements. 
Note that teleportation within path superpositions will be a superposition of quantum processes.

In some architectures for the Quantum Internet, the complex entanglement swapping, fidelity measurements, retransmission (if required), and entanglement distribution is assigned to processes occurring directly  in  the Application and/or Transport layers, perhaps coupled to a software defined networking (SDN) architecture. In this regard, we note the architecture proposed in~\cite{contimal} where satellites control the entanglement distribution on Earth directly from space. With  quantum memory in place, we can envisage this type of architecture widely distributing entanglement to many pairs of ground stations, which can then re-distribute the entanglement via direct transmission to those stations fiber linked within $\sim100$~km. In this architecture, the application layer requests a teleportation channel, and the network checks if available, and if not, creates, on-demand, the teleportation channel through direct transmission from space coupled with local direct transmissions.


\section{Discussion}
Our discussion above has developed a hybrid classical-quantum header for next-generation quantum networks. We have described how current IPv6 frameworks can encapsulate a surprisingly wide range of possible quantum processes of import at the IP layer. By slowly enhancing routers with the appropriate software and hardware, a pathway to the inclusion of quantum-only effects can be added to existing infrastructure. The architecture proposed is backward compatible with existing network hardware, as routers ill-equipped for quantum processing will simply ignore the Quantum Extension Headers.

We have not seriously considered the role to be played by a quantum software defined network (QSDN) arrangement~\cite{qsdn}. Although complex, this form of quantum software-based solution leads in some circumstances to new network functionality over the use of hybrid classical-quantum headers discussed here. Given that QSDN could play a role in many functions, such as entanglement swapping, and QKD enhanced security; going forward we believe it possible that quantum networks may well operate based on some form of quantum enabled IPv6 Extension Headers and QSDN. 

We note that it is also possible to have some header information in the form of quantum states~\cite{cal8}. This again would likely require the intervention of a QSDN to assist controlling information transfer as that quantum information could not be read by all routers on route. 

Finally, beyond the many advantages to communication systems we have previously noted, the wider quantum-enabled functionality  allowed by our proposed framework should allow for new or improved applications demonstrating a quantum advantage.   We refer the reader to recent reviews that present many new applications for quantum networks that the framework presented here could potentially assist with\cite{review1, review2, review2b, review3}. We believe that these exciting new quantum-enabled networks will provide opportunities for scenarios and applications that have not yet been discovered.

\section{Conclusion}\label{sec:conclude}
The Global Quantum Internet is emerging, but much work is still required to finalize the rules and protocols for its operation. Many interesting works attempting to set these rules abound, ranging from traditional layered approaches, to radical new designs which completely abandon the architecture on which the Classical Internet is built. In this work, we highlighted how a wide range of quantum-only processes can be accommodated at the IP layer via some modifications to the IPv6 Extension Header model. 

Our framework allows for future `simple' functionality such as direct transfer of a quantum state, the teleportation of a quantum state, and the superposition of two routes. Beyond this, a plethora of complexity is also allowed for, such as web-like superpositions of many routes extending over the entire network, and even superpositions of different quantum processes at different routers in the network. This allowed complexity provides for significant future proofing of the proposed architecture as these exciting new networks develop. 

The framework we have outlined requires further  detail and consideration. Future community involvement and a rigorous standardization process will be needed before the concepts proposed here or elsewhere can be put into widespread practice. 
\section*{\label{sec:Acknowledgments}Acknowledgments}
The author thanks Professors Angela Sara Cacciapuoti and Marcello Caleffi for their valuable comments on this work. The author also thanks the  University of Naples Federico II for hosting a sabbatical visit during which this work was conducted.

\bibliographystyle{IEEEtran}

\bibliography{mybib}

@INPROCEEDINGS{shi6,
  author={Shi, Wenbo and Kundu, Neel Kanth and McKay, Matthew R. and Malaney, Robert},
  booktitle={2025 International Conference on Quantum Communications, Networking, and Computing (QCNC)}, 
  title={Error-Mitigated Quantum Random Access Memory}, 
  year={2025},
  volume={},
  number={},
  pages={32-39},
  doi={10.1109/QCNC64685.2025.00015},
url={https://doi.org/10.1109/QCNC64685.2025.00015}}

@article{optmem,
  title = {Fiber-integrated quantum memory for telecom light},
  author = {Bonsma-Fisher, K. A. G. and Hnatovsky, C. and Grobnic, D. and Mihailov, S. J. and Bustard, P. J. and England, D. G. and Sussman, B. J.},
  journal = {Phys. Rev. A},
  volume = {108},
  issue = {1},
  pages = {012606},
  numpages = {8},
  year = {2023},
  month = {Jul},
  publisher = {American Physical Society},
  doi = {10.1103/PhysRevA.108.012606},
  url = {https://link.aps.org/doi/10.1103/PhysRevA.108.012606}
}

@article{tel2,
  title     = {Progress in quantum teleportation},
  author    = {Hu, Xiao-Min and Guo, Yu and Liu, Bi-Heng and Li, Chuan-Feng and Guo, Guang-Can},
  journal   = {Nature Reviews Physics},
  year      = {2023},
  volume    = {5},
  number    = {6},
  pages     = {339--353},
  doi       = {10.1038/s42254-023-00588-x},
  url       = {https://doi.org/10.1038/s42254-023-00588-x},
  publisher = {Springer Science and Business Media LLC}
}

@article{tel1,
  title = {Teleporting an unknown quantum state via dual classical and Einstein-Podolsky-Rosen channels},
  author = {Bennett, Charles H. and Brassard, Gilles and Crepeau, Claude and Jozsa, Richard and Peres, Asher and Wootters, William K.},
  journal = {Phys. Rev. Lett.},
  volume = {70},
  issue = {13},
  pages = {1895--1899},
  numpages = {0},
  year = {1993},
  month = {Mar},
  publisher = {American Physical Society},
  doi = {10.1103/PhysRevLett.70.1895},
  url = {https://link.aps.org/doi/10.1103/PhysRevLett.70.1895}
}

@article{Oi,
  title = {Interference of Quantum Channels},
  author = {Oi, Daniel K. L.},
  journal = {Phys. Rev. Lett.},
  volume = {91},
  issue = {6},
  pages = {067902},
  numpages = {4},
  year = {2003},
  publisher = {American Physical Society},
  doi = {10.1103/PhysRevLett.91.067902},
  url = {https://link.aps.org/doi/10.1103/PhysRevLett.91.067902}
}

@ARTICLE{wenbo1,
  author={Shi, Wenbo and Malaney, Robert},
  journal={IEEE Network}, 
  title={Quantum Routing for Emerging Quantum Networks}, 
  year={2024},
  volume={38},
  number={1},
  pages={140-146},
  doi={10.1109/MNET.2023.3317821},
  url={https://doi.org/10.1109/MNET.2023.3317821}}

@article{callno1,
  title={Quantum Internet protocol stack: A comprehensive survey},
  author={Illiano, Jessica and Caleffi, Marcello and Manzalini, Antonio and Cacciapuoti, Angela Sara},
  journal={Computer Networks},
  volume={213},
  pages={109092},
  year={2022},
  publisher={Elsevier},
  doi={10.1016/j.comnet.2022.109092},
  url={https://doi.org/10.1016/j.comnet.2022.109092}}

@article{y1,
	doi = {10.1088/1367-2630/ab05f7},
	url = {https://doi.org/10.1088/1367-2630/ab05f7},
	year = 2019,
	publisher = {IOP Publishing},
	volume = {21},
	number = {3},
	pages = {033003},
	author = {A Pirker and W Dür},
	title = {A quantum network stack and protocols for reliable entanglement-based networks},
	journal = {New Journal of Physics}
}

@article{y2,
  author    = {Rodney Van Meter and Thaddeus D. Ladd and William J. Munro and Kae Nemoto},
  title     = {System Design for a Long-Line Quantum Repeater},
  journal   = {IEEE/ACM Transactions on Networking},
  volume    = {17},
  number    = {3},
  pages     = {1002--1013},
  year      = {2009},
  doi       = {10.1109/TNET.2008.927260},
   url       = {https://doi.org/10.1109/TNET.2008.927260},
  eprint    = {0705.4128},
  archivePrefix = {arXiv},
  primaryClass = {quant-ph}
}

@inproceedings{y3,
  title={A link layer protocol for quantum networks},
  author={Dahlberg, Axel and Skrzypczyk, Matthew and Cavalcanti, David and Wehner, Stephanie},
  booktitle={Proceedings of the ACM Special Interest Group on Data Communication (SIGCOMM)},
  pages={159--173},
  year={2019},
  doi={10.1145/3341302.3342070},
  url={https://dl.acm.org/doi/10.1145/3341302.3342070}
}

@inproceedings{y4,
author = {Kozlowski, Wojciech and Dahlberg, Axel and Wehner, Stephanie},
title = {Designing a quantum network protocol},
year = {2020},
isbn = {9781450379489},
publisher = {{ACM}},
address = {NY, USA},
url = {https://doi.org/10.1145/3386367.3431293},
doi = {10.1145/3386367.3431293},

booktitle = {Proceedings of the 16th International Conference on Emerging Networking Experiments and Technologies},
pages = {1–16},
numpages = {16},
location = {Barcelona, Spain}
}

@article{Durno1,
  title = {A resource-centric, task-based approach to quantum network control},
  author = {Pirker, Alexander and Munoz, Belen and Dür, Wolfgang},
  journal = {arXiv preprint arXiv:2507.12030},
  year = {2025},
  url = {https://arxiv.org/abs/2507.12030}
}

@article{callno2,
  title     = {A Quantum Internet Protocol Suite Beyond Layering},
  author    = {Cacciapuoti, Angela Sara and Caleffi, Marcello},
  journal   = {IEEE Transactions on Network Science and Engineering},
  year      = {2026},
  publisher = {IEEE},
  doi       = {10.1109/TNSE.2026.3679795},
  url       = {https://doi.org/10.1109/TNSE.2026.3679795}
}

@article{qramo,
  title = {Quantum Random Access Memory},
  author = {Giovannetti, Vittorio and Lloyd, Seth and Maccone, Lorenzo},
  journal = {Phys. Rev. Lett.},
  volume = {100},
  issue = {16},
  pages = {160501},
  numpages = {4},
  year = {2008},
  publisher = {American Physical Society},
  doi = {10.1103/PhysRevLett.100.160501},
  url = {https://link.aps.org/doi/10.1103/PhysRevLett.100.160501}
}

@ARTICLE{contimal,
  author={Conti, Andrea and Malaney, Robert and Win, Moe Z.},
  journal={IEEE Communications Magazine}, 
  title={Satellite-Terrestrial Quantum Networks and the Global Quantum Internet}, 
  year={2024},
  volume={62},
  number={10},
  pages={34-39},
  doi={10.1109/MCOM.007.2300854},
url={https://doi.org/10.1109/MCOM.007.2300854}}

@ARTICLE{malaney5,
  author={Al-Rabayah, Mohammad and Malaney, Robert},
  journal={IEEE Transactions on Vehicular Technology}, 
  title={A New Scalable Hybrid Routing Protocol for {VANETs}}, 
  year={2012},
  volume={61},
  number={6},
  pages={2625-2635},
  doi={10.1109/TVT.2012.2198837},
 url={https://doi.org/10.1109/TVT.2012.2198837}}

@article{18p4,
  title={Experimental Demonstration of Datagram Switching With Monitoring in Quantum Wrapper Networks},
  author={On, Mehmet Berkay and Proietti, Roberto and G{\"u}l, Gamze and Kanter, Gregory S. and Singh, Sandeep Kumar and Kumar, Prem and Yoo, S. J. Ben},
  journal={Journal of Lightwave Technology},
  volume={42},
  number={10},
  pages={3504--3517},
  year={2024},
  publisher={IEEE},
  doi={10.1109/JLT.2024.3362292},
  url={https://doi.org/10.1109/JLT.2024.3362292}
}

@INPROCEEDINGS{class34,
  author={Hendriks, Luuk and Velan, Petr and Schmidt, Ricardo de O. and de Boer, Pieter-Tjerk and Pras, Aiko},
  booktitle={2017 Network Traffic Measurement and Analysis Conference (TMA)}, 
  title={Threats and surprises behind IPv6 extension headers}, 
  year={2017},
  volume={},
  number={},
  pages={1-9},
  doi={10.23919/TMA.2017.8002912},
  url={https://doi.org/10.23919/TMA.2017.8002912}
  }

@inproceedings{class33,
  title={Facilitating Data Usage Control Through {IPv6} Extension Headers},
  author={Qarawlus, Haydar and Hellmeier, Malte and Howar, Falk},
  booktitle={Proceedings of the 14th International Conference on Data Management Technologies and Applications (DATA)},
  pages={526--535},
  year={2025},
  organization={SciTePress},
  doi={10.5220/0013566200003967},
    url={https://doi.org/10.5220/0013566200003967}
}

@misc{illi,
      title={Quantum Routers: A Switching-Fabric Framework for Quantum-Native Forwarding}, 
      author={Jessica Illiano and Caterina De Risi and Angela Sara Cacciapuoti and Marcello Caleffi},
      year={2026},
      eprint={2606.17773},
      archivePrefix={arXiv},
      primaryClass={quant-ph},
      url={https://arxiv.org/abs/2606.17773}, 
}

@article{18p3,
  author = {Yoo, S. J. Ben and On, Mehmet Berkay and Gul, Gamze and Proietti, Roberto and Kanter, Gregory S. and Kumar, Prem},
  title = {Quantum Wrapper Networking},
  journal = {IEEE Communications Magazine},
  year = {2024},
  volume = {62},
  number = {3},
  pages = {148-155},
  doi = {10.1109/MCOM.001.2300067},
  url = {https://ieeexplore.ieee.org/document/9593032/}
}

@Article{18p2,
author={Mandil, Reem
and DiAdamo, Stephen
and Qi, Bing
and Shabani, Alireza},
title={Quantum key distribution in a packet-switched network},
journal={npj Quantum Information},
year={2023},
month={Sep},
day={09},
volume={9},
number={1},
pages={85},
issn={2056-6387},
doi={10.1038/s41534-023-00757-x},
url={https://doi.org/10.1038/s41534-023-00757-x}
}

@article{18p1,
  title = {Packet switching in quantum networks: A path to the quantum internet},
  author = {DiAdamo, Stephen and Dahlberg, Axel and Skrzypczyk, Matthew and Wehner, Stephanie},
  journal = {Phys. Rev. Res.},
  volume = {4},
  issue = {4},
  pages = {043064},
  year = {2022},
  publisher = {American Physical Society},
  doi = {10.1103/PhysRevResearch.4.043064},
  url = {https://link.aps.org/doi/10.1103/PhysRevResearch.4.043064}
}

@misc{allproc,
      title={Path superposition as resource for perfect quantum teleportation with separable states}, 
      author={Sayan Mondal and Priya Ghosh and Ujjwal Sen},
      year={2025},
      eprint={2505.11398},
      archivePrefix={arXiv},
      primaryClass={quant-ph},
      url={https://arxiv.org/abs/2505.11398}, 
}

@ARTICLE{osi,
  author={Zimmermann, H.},
  journal={IEEE Transactions on Communications}, 
  title={{OSI} Reference Model - The {ISO} Model of Architecture for Open Systems Interconnection}, 
  year={1980},
  volume={28},
  number={4},
  pages={425-432},
  doi={10.1109/TCOM.1980.1094702},
url={https://doi.org/10.1109/TCOM.1980.1094702}}

@ARTICLE{cal8,
  author={Cacciapuoti, Angela Sara and Illiano, Jessica and Caleffi, Marcello},
  journal={IEEE Network}, 
  title={Quantum Internet Addressing}, 
  year={2024},
  volume={38},
  number={1},
  pages={104-111},
  doi={10.1109/MNET.2023.3328393},
url={https://doi.org/10.1109/MNET.2023.3328393}}

@ARTICLE{caleffi1,
  author={Caleffi, Marcello and Cacciapuoti, Angela Sara},
  journal={IEEE Transactions on Communications}, 
  title={Quantum Internet Architecture: Unlocking Quantum-Native Routing via Quantum Addressing}, 
  year={2026},
  volume={74},
  number={},
  pages={3577-3599},
  doi={10.1109/TCOMM.2025.3650397},
url={https://doi.org/10.1109/TCOMM.2025.3650397}}

@techreport{rfc8200,
  author = {S. Deering and R. Hinden},
  title = {Internet Protocol, Version 6 ({IPv6}) Specification},
  howpublished = {Internet Requests for Comments},
  type = {RFC},
  number = 8200,
  year = 2017,
  issn = {2070-1721},
  publisher = {RFC Editor},
  institution = {RFC Editor},
  doi = {10.17487/RFC8200},
  url = {https://www.rfc-editor.org/info/rfc8200}
}

@article{web3,
  title = {Improved routing strategies for Internet traffic delivery},
  author = {Echenique, Pablo and G\'omez-Garde\~nes, Jes\'us and Moreno, Yamir},
  journal = {Phys. Rev. E},
  volume = {70},
  issue = {5},
  pages = {056105},
  numpages = {4},
  year = {2004},
  month = {Nov},
  publisher = {American Physical Society},
  doi = {10.1103/PhysRevE.70.056105},
  url = {https://link.aps.org/doi/10.1103/PhysRevE.70.056105}
}

@ARTICLE{qsdn,
  author={Aguado, Alejandro and Lopez, Victor and Lopez, Diego and Peev, Momtchil and Poppe, Andreas and Pastor, Antonio and Folgueira, Jesus and Martin, Vicente},
  journal={IEEE Communications Magazine}, 
  title={The Engineering of Software-Defined Quantum Key Distribution Networks}, 
  year={2019},
  volume={57},
  number={7},
  pages={20-26},
  doi={10.1109/MCOM.2019.1800763},
  url={https://doi.org/10.1109/MCOM.2019.1800763}}

@ARTICLE{infin,
author={Kristj{\'a}nsson, Hl{\'e}r
and Zhong, Yan
and Munson, Anthony
and Chiribella, Giulio},
title={Quantum networks with coherent routing of information through multiple nodes},
journal={npj Quantum Information},
year={2024},
day={20},
volume={10},
number={1},
pages={131},
issn={2056-6387},
doi={10.1038/s41534-024-00919-5},
url={https://doi.org/10.1038/s41534-024-00919-5}}

@article{port,
  title = {Asymptotic Teleportation Scheme as a Universal Programmable Quantum Processor},
  author = {Ishizaka, Satoshi and Hiroshima, Tohya},
  journal = {Physical Review Letters},
  volume = {101},
  issue = {24},
  pages = {240501},
  numpages = {4},
  year = {2008},
  publisher = {American Physical Society},
  doi = {10.1103/PhysRevLett.101.240501},
  url = {https://link.aps.org/doi/10.1103/PhysRevLett.101.240501}
}

@misc{iana,
  author       = {{Internet Assigned Numbers Authority (IANA)}},
  title        = {{Internet Protocol Version 6 ({IPv6}) Parameters}},
  howpublished = {\url{https://www.iana.org/assignments/ipv6-parameters/ipv6-parameters.xhtml}},
  year         = {2025},
  note         = {Last updated: 2025-12-24}
}

@article{QRAM,
  doi = {10.22331/q-2025-12-02-1922},
  url = {https://doi.org/10.22331/q-2025-12-02-1922},
  title = {{QRAM}: {A} {S}urvey and {C}ritique},
  author = {Jaques, Samuel and Rattew, Arthur G.},
  journal = {{Quantum}},
  issn = {2521-327X},
  publisher = {{Verein zur F{\"{o}}rderung des Open Access Publizierens in den Quantenwissenschaften}},
  volume = {9},
  pages = {1922},
  year = {2025}
}

@article{DelSanto,
  title = {Two-Way Communication with a Single Quantum Particle},
  author = {Del Santo, Flavio and Daki\ifmmode \acute{c}\else \'{c}\fi{}, Borivoje},
  journal = {Physical Review Letters},
  volume = {120},
  issue = {6},
  pages = {060503},
  numpages = {6},
  year = {2018},
  publisher = {American Physical Society},
  doi = {10.1103/PhysRevLett.120.060503},
  url = {https://link.aps.org/doi/10.1103/PhysRevLett.120.060503}
}

@article{Lemr2013,
  title = {Resource-efficient linear-optical quantum router},
  author = {Lemr, Karel and Bartkiewicz, Karol and \v{C}ernoch, Anton\'{\i}n and Soubusta, Jan},
  journal = {Phys. Rev. A},
  volume = {87},
  issue = {6},
  pages = {062333},
  numpages = {7},
  year = {2013},
  publisher = {American Physical Society},
  doi = {10.1103/PhysRevA.87.062333},
  url ={https://doi.org/10.1103/PhysRevA.87.062333}
  


}

@article{Gisin,
  title = {Error Filtration and Entanglement Purification for Quantum Communication},
  author = {Gisin, Nicolas and Linden, Noah and Massar, Serge and Popescu, Sandu},
  journal = {Physical Review A},
  volume = {72},
  issue = {1},
  pages = {012338},
  numpages = {13},
  year = {2005},
  publisher = {American Physical Society},
  doi = {10.1103/PhysRevA.72.012338},
  url = {https://link.aps.org/doi/10.1103/PhysRevA.72.012338}
}

@article{Chiri,
	doi = {10.1098/rspa.2018.0903},
	url = {https://doi.org/10.1098/rspa.2018.0903},
	year = 2019,
	publisher = {The Royal Society},
	volume = {475},
	number = {2225},
	pages = {20180903},
	author = {Giulio Chiribella and Hlér Kristjánsson},
	title = {Quantum {S}hannon theory with superpositions of trajectories},
	journal = {Proceedings of the Royal Society A: Mathematical, Physical and Engineering Sciences}
}

@article{casor,
  title = {Enhanced Communication with the Assistance of Indefinite Causal Order},
  author = {Ebler, Daniel and Salek, Sina and Chiribella, Giulio},
  journal = {Phys. Rev. Lett.},
  volume = {120},
  issue = {12},
  pages = {120502},
  numpages = {5},
  year = {2018},
  month = {Mar},
  publisher = {American Physical Society},
  doi = {10.1103/PhysRevLett.120.120502},
  url = {https://link.aps.org/doi/10.1103/PhysRevLett.120.120502}
}

@article{sen25,
  title={To share and not share a singlet: control qubit and nonclassicality in teleportation},
  author={Sen, Kornikar and Ajith, Adithi and Halder, Saronath and Sen, Ujjwal},
  journal={Journal of Physics A: Mathematical and Theoretical},
  volume={58},
  number={5},
  pages={055302},
  year={2025},
  publisher={IOP Publishing},
  doi={10.1088/1751-8121/ada64a},
  url = {https://doi.org/10.1088/1751-8121/ada64a}
}

@ARTICLE{review2,
  author={Singh, Amoldeep and Dev, Kapal and Siljak, Harun and Joshi, Hem Dutt and Magarini, Maurizio},
  journal={IEEE Communications Surveys and Tutorials}, 
  title={Quantum Internet—Applications, Functionalities, Enabling Technologies, Challenges, and Research Directions}, 
  year={2021},
  volume={23},
  number={4},
  pages={2218-2247},
  doi={10.1109/COMST.2021.3109944},
url={https://doi.org/10.1109/COMST.2021.3109944}
}

@article{review1,
  title = {Advances in Quantum Cryptography},
  author = {Pirandola, S. and Andersen, U. L. and Banchi, L. and Berta, M. and Bunandar, D. and Colbeck, R. and Englund, D. and Gehring, T. and Lupo, C. and Ottaviani, C. and Pereira, J. L. and Razavi, M. and Shaari, J. S. and Tomamichel, M. and Usenko, V. C. and Vallone, G. and Villoresi, P. and Wallden, P.},
  journal = {Advances in Optics and Photonics},
  volume = {12},
  number = {4},
  pages = {1012--1236},
  year = {2020},
  publisher = {Optica Publishing Group},
  doi = {10.1364/AOP.361502},
  url = {https://opg.optica.org/aop/abstract.cfm?uri=aop-12-4-1012}
}

@article{review2b,
author = {Rozenman, Georgi Gary and Kundu, Neel Kanth and Liu, Ruiqi and Zhang, Leyi and Maslennikov, Alona and Reches, Yuval and Youm, Heung Youl},
title = {The quantum internet: A synergy of quantum information technologies and 6G networks},
journal = {IET Quantum Communication},
volume = {4},
number = {4},
pages = {147-166},
doi = {https://doi.org/10.1049/qtc2.12069},
url = {https://ietresearch.onlinelibrary.wiley.com/doi/abs/10.1049/qtc2.12069},
eprint = {https://ietresearch.onlinelibrary.wiley.com/doi/pdf/10.1049/qtc2.12069},
year = {2023}
}

@article{review3,
  title = {Continuous-variable quantum communication},
  author = {Usenko, Vladyslav C. and Ac\'{\i}n, Antonio and All\'eaume, Romain and Andersen, Ulrik L. and Diamanti, Eleni and Gehring, Tobias and Hajomer, Adnan A. E. and Kanitschar, Florian and Pacher, Christoph and Pirandola, Stefano and Pruneri, Valerio},
  journal = {Rev. Mod. Phys.},
  volume = {98},
  issue = {1},
  pages = {015003},
  numpages = {58},
  year = {2026},
  publisher = {American Physical Society},
  doi = {10.1103/mgj7-t6d3},
  url = {https://link.aps.org/doi/10.1103/mgj7-t6d3}
}

@article{laserpack,
author = {Yichi Zhang  and Robert Broberg  and Alan Zhu  and Gushu Li  and Li Ge  and Jonathan M. Smith  and Liang Feng },
title = {Classical-decisive quantum internet by integrated photonics},
journal = {Science},
volume = {389},
number = {6763},
pages = {940-944},
year = {2025},
doi = {10.1126/science.adx6176},
URL = {https://www.science.org/doi/abs/10.1126/science.adx6176},
eprint = {https://www.science.org/doi/pdf/10.1126/science.adx6176}}

@Article{link1,
author={Pompili, M.
and Delle Donne, C.
and te Raa, I.
and van der Vecht, B.
and Skrzypczyk, M.
and Ferreira, G.
and de Kluijver, L.
and Stolk, A. J.
and Hermans, S. L. N.
and Pawe{\l}czak, P.
and Kozlowski, W.
and Hanson, R.
and Wehner, S.},
title={Experimental demonstration of entanglement delivery using a quantum network stack},
journal={npj Quantum Information},
year={2022},
volume={8},
number={1},
pages={121},
issn={2056-6387},
doi={10.1038/s41534-022-00631-2},
url={https://doi.org/10.1038/s41534-022-00631-2}
}

@ARTICLE{link2,
  author={Baier, Benedikt and Rosenauer, Ria and Li, Vili and Deppe, Christian and Kellerer, Wolfgang},
  journal={IEEE Transactions on Quantum Engineering}, 
  title={Combined Physical- and Link-Layer Protocols for Quantum Networks}, 
  year={2026},
  volume={7},
  number={},
  pages={1-15},
  doi={10.1109/TQE.2025.3630201},
url={https://doi.org/10.1109/TQE.2025.3630201}}

@ARTICLE{internetv1,
  author={Cerf, V. and Kahn, R.},
  journal={IEEE Transactions on Communications}, 
  title={A Protocol for Packet Network Intercommunication}, 
  year={1974},
  volume={22},
  number={5},
  pages={637-648},
  doi={10.1109/TCOM.1974.1092259},
url={https://doi.org/10.1109/TCOM.1974.1092259}}

@article{ref3,
  title = {Micius quantum experiments in space},
  author = {Lu, Chao-Yang and Cao, Yuan and Peng, Cheng-Zhi and Pan, Jian-Wei},
  journal = {Rev. Mod. Phys.},
  volume = {94},
  issue = {3},
  pages = {035001},
  numpages = {46},
  year = {2022},
  publisher = {American Physical Society},
  doi = {10.1103/RevModPhys.94.035001},
  url = {https://link.aps.org/doi/10.1103/RevModPhys.94.035001}
}

\end{document}